\documentclass[prb,reprint,amsmath,amssymb,showpacs,twocolumn]{revtex4-1}

\usepackage[dvipdfmx]{graphicx}
\usepackage{bm}
\usepackage{mathrsfs}
\usepackage{color}

\DeclareMathOperator{\trace}{Tr}

\DeclareMathOperator{\sgn}{sgn}

\begin{document}
\title{Fragile surface zero-energy flat bands in three-dimensional chiral superconductors}
\author{Shingo Kobayashi$^1$}
\author{Yukio Tanaka$^1$}
\author{Masatoshi Sato$^2$}
\affiliation{$^1$Department of Applied Physics, Nagoya University, Nagoya 464-8603, Japan \\ $^2$Yukawa Institute for Theoretical Physics, Kyoto University, Kyoto 606-8502, Japan}

\date{\today}

\begin{abstract}
We study surface zero-energy flat bands in three-dimensional chiral superconductors with $p_z (p_x +ip_y)^{\nu}$-wave pairing symmetry ($\nu$ is a nonzero integer), based on topological arguments and tunneling conductance. It is shown that the surface flat bands are fragile against (i) the surface misorientation and (ii) the surface Rashba spin-orbit interaction. The fragility of (i) is specific to chiral SCs, whereas that of (ii) happens for general odd-parity SCs. We demonstrate that these flat band instabilities vanish or suppress a zero-bias conductance peak in a normal/insulator/superconductor junction, which behavior is clearly different from high-$T_c$ cuprates and noncentrosymmetric superconductors. By calculating the angle resolved conductance, we also discuss a topological surface state associated with the coexistence of line and point nodes. 
\end{abstract}
\maketitle
\section{Introduction}
Gapless phases of matter have received enormous attention in recent years. In the context of unconventional superconductors (SCs), such gapless phases manifest as nodal excitations in superconducting gaps, and they have played important roles in determination of the pairing symmetry. Recent developments on topological classification~\cite{Volovik:2003,Horava:2005,Zhao:2013,Sato:2006,Beri:2010,SAYang:2014,Shiozaki:2014,Kobayashi:2014,Chiu:2014}   have deepened the understanding of the nodal stability. From the topological perspective, the stability of the nodal structure is ensured by topological numbers, which predict surface zero energy Andreev bound states at the same time, due to the bulk-boundary correspondence. In particular, line nodes induce surface zero energy flat bands in projected surface Brillouin zone (BZ)~\cite{Sato:2011,Tanaka:2012,Schnyder:2011,Matsuura:2013,Volovik:2013}. The existence of the surface zero-energy flat bands results in a zero-bias conductance peak (ZBCP) in the tunneling spectroscopy~\cite{Hu:1994,Matsumoto:1995,Tanaka:1995,Kashiwaya:2000,Lofwander:2001}, which has been observed in high-T$_c$ cuprates~\cite{Kashiwaya:1998,Covington:1997,Alff:1997,Wei:1998,Iguchi:2000,Biswas:2002,Chesca:2008}. In addition, similar ZBCPs have been anticipated theoretically in time-reversal invariant (TRI) noncentrosymmetric SCs~\cite{Iniotakis:2007,Tanaka:2010,Brydon:2011,Schnyder:2012,Yada:2011} such as CePt$_3$Si~\cite{Bauer:2004,Izawa:2005,Bonalde:2005}, CeRh$_3$Si~\cite{Kimura:2005}, and CeIrSi$_3$~\cite{Sugitani:2006}. 

\begin{figure}[tbp]
\centering
 \includegraphics[width=8cm]{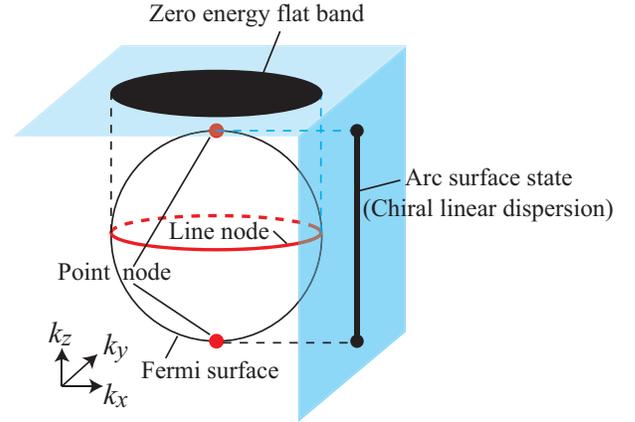}
  \caption{(color online) Schematic picture of topological surface states in 3D chiral SCs with $\nu =1$, where a spherical Fermi surface is assumed. The chiral SC hosts both line and point nodes, which induces a zero-energy flat band and an arc surface state (chiral linear dispersion) at each projected surface, respectively. } \label{fig:chiral-topology}
 \end{figure}

Up to this time, line node in TRI SCs have been considered. In TRI SCs, there is chiral symmetry that is obtained as a combination of particle-hole (PHS) and time-reversal symmetries (TRS) in most topological arguments. The chiral symmetry makes it possible to define a one-dimensional (1D) winding number, which ensures the stability of line nodes and the existence of surface flat bands. However, for heavy fermion SCs such as UPt$_3$~\cite{Sauls:1994,Joynt:2002,Schemm:2014} and URu$_2$Si$_2$~\cite{Kasahara:2007,Shibauchi:2014,Schemm:2015}, line nodes in time-reversal breaking gap functions have been also proposed. These materials are candidates of 3D chiral superconductors with $p_z(p_x+ip_y)^\nu$ pairing symmetry. $\nu=1$ corresponds to chiral d-wave pairing　(URu$_2$Si$_2$), and $\nu=2$ to chiral f-wave pairing (B-phase of UPt$_3$). These gap functions support a horizontal line node on the $k_x k_y$ plane, and point nodes in the $k_z$-axis. Interestingly, it has been shown that the 3D chiral SCs support zero energy surface flat bands on a surface perpendicular to the $z$-axis~\cite{Goswami:2013,Goswami:2014}. (See Fig.~\ref{fig:chiral-topology}). 

 In this paper, we address the stability of the line node and the surface flat band in the 3D chiral SCs, based on topological arguments and tunneling conductance. Since the chiral SCs break time-reversal symmetry, the forementioned chiral symmetry is absent. Nevertheless, by using a momentum-dependent gauge symmetry, we find a momentum-dependent chiral symmetry specific to the chiral SCs, which allows us to define a similar 1D winding number. This winding number explains the existence of zero energy surface flat bands on a surface normal to the $z$-axis. In contrast to the original chiral symmetry, the model-dependent chiral symmetry is fragile, and the surface zero energy flat bands disappear when a surface is not normal to the $z$-axis. We find that in the latter situation, arc surface states appear, instead. (See Fig.~\ref{fig:chiral-angle}). The surface arcs connect the projections of the point nodes and the line nodes on the surface BZ. We also reveal the topological origin of the arc structures. The main purpose of this paper is to clarify the disappearance of zero energy flat bands and explore the interplay between surface states originating from line and point nodes in the higher order chiral SCs.

We also argue the parity-dependence of the stability. For odd $\nu$, the parity of the gap function is even, while for even $\nu$, the parity is odd. In our previous paper~\cite{Kobayashi:2014}, we have uncovered that line nodes in odd-parity SCs are topologically unstable, and the corresponding surface states are fragile against the surface Rashba spin-orbit interaction (RSOI). Taking into account the RSOI in the calculation of tunneling conductance, we confirm the flat band instability in the even $\nu$ case as the suppression of the ZBCP with increasing the magnitude of the RSOI.

This paper is organized as follows. In Sec.~\ref{sec:model}, we discuss our model and symmetries that the 3D chiral SCs host. In Sec.~\ref{sec:top}, 1D and 2D topological numbers in 3D chiral SCs are introduced and their relation to zero-energy Andreev bound states are discussed in subsection~\ref{subsec:top-surface}.  In Sec.~\ref{sec:conductance}, we show tunneling conductance in a normal/insulator/superconductor junction for chiral SCs. The influence of the misorientation angle and the RSOI on the conductance is discussed in subsection~\ref{subsec:misorientation} and ~\ref{subsec:Rashba}, respectively. Finally, we summarize this paper in Sec.~\ref{sec:sum}.

\section{Model systems}
\label{sec:model}
We phenomenologically model 3D chiral SCs with $k_z (k_x + ik_y)^{\nu}$-wave pairing symmetry ($\nu=0,1,2, \cdots$) as a single band system described by the Bogoliubov-de Gennes (BdG) Hamiltonian $\mathcal{H} = \frac{1}{2} \sum_{\bm{k}} \Psi_{\bm{k}}^{\dagger} H(\bm{k}) \Psi_{\bm{k}}$, with $\Psi_{\bm{k}}^T = (c_{\bm{k},\uparrow},c_{\bm{k},\downarrow},c_{-\bm{k},\uparrow}^{\dagger},c_{-\bm{k},\downarrow}^{\dagger} )$ and
\begin{align}
H(\bm{k}) = \begin{pmatrix} \mathcal{E} (\bm{k}) & \Delta (\bm{k}) \\ \Delta (\bm{k})^{\dagger} & -\mathcal{E} ^T(-\bm{k})\end{pmatrix}. \label{eq:BdG}
\end{align}   
Here $\mathcal{E} (\bm{k}) $ describes a normal dispersion with TRS and $\Delta (\bm{k})$ takes 
 \begin{align}
  \Delta (\bm{k}) = \begin{cases} \frac{\Delta_0}{k_F^{\nu +1}} k_z (k_x + ik_y)^{\nu} i s_y &\; \nu: {\rm odd}, \\  \frac{\Delta_0}{k_F^{\nu+1}} k_z (k_x + ik_y)^\nu s_x &\; \nu: {\rm even}.\end{cases} \label{eq:gapfn}
 \end{align}
The parity of the gap function is even (odd) when $\nu$ is odd (even). $k_F$ is the Fermi wavelength and the direction of $\bm{d}$-vector for the odd parity gap function is chosen to be parallel to the $z$ direction for convenience. Note that the $k_z (k_x + ik_y)^{\nu}$-wave pairing symmetry is realized when the gap function respects a 2D irreducible representation of point groups; for example, UPt$_3$ and URu$_2$Si$_2$ correspond to the $E_{2u}$ representation of $D_{6h}$ and the $E_g$ representation of $D_{4h}$, respectively.   
 
 The BdG Hamiltonian~(\ref{eq:BdG}) hosts several discrete symmetries that are relevant to topological numbers discussed in the next section. First of all, the BdG Hamiltonian hosts PHS: $CH(\bm{k}) C^{-1} = -H(-\bm{k})$ with $C= s_0 \tau_x K$, where $s_i =(1,{\bm \sigma})$ and $\tau_i = (1,{\bm \tau})$ are the Pauli matrices in spin and Nambu spaces, respectively, and $K$ is the complex conjugation. For $\nu =0$, the BdG Hamiltonian also supports TRS: $TH(\bm{k})T^{-1} = H(-\bm{k})$, where $T=i s_y \tau_0 K$. We also impose inversion symmetry on the present model such that $PH(\bm{k})P^{-1} = H(-\bm{k})$ with $P=s_0 \tau_0$ for even parity and $P= s_0 \tau_z$ for odd parity.
  
Besides these fundamental symmetries, the system with nonzero $\nu$ also has an accidental symmetry, which we call pseudo TRS. The gap function $k_z (k_x + ik_y)^{\nu}$ is rewritten as $k_z(k_x^2+k_y^2)^{\frac{\nu}{2}} e^{i \nu \varphi_{\bm{k}}}$ with $\varphi_{\bm{k}} = \tan^{-1}\frac{k_y}{k_x}$. Thus, by the local gauge transformation, $U_{\varphi_{\bm{k}}} H(\bm{k}) U_{\varphi_{\bm{k}}}^{\dagger}$ with $U_{\varphi_{\bm{k}}} = e^{- i \frac{\nu }{2} \varphi_{\bm{k}} s_0 \tau_z}$, the gap function becomes $\frac{\Delta_0}{k_F^{\nu +1}} k_z(k_x^2+k_y^2)^{\frac{\nu}{2}}$, so the system recovers TRS. In other words, the BdG Hamiltonian~(\ref{eq:BdG}) has the following momentum dependent pseudo TRS.
 \begin{align}
  U_{\varphi_{\bm{k}}}^{\dagger} TU_{\varphi_{\bm{k}}} H(\bm{k})U_{\varphi_{\bm{k}}}^{\dagger} T^{\dagger} U_{\varphi_{\bm{k}}}= H(-\bm{k}), \label{eq:pseudo-TRS}
 \end{align} 
where $U_{\varphi_{\bm{k}}} = s_0 \tau_0$ when $\varphi_{\bm{k}} = 0$. As we will discuss in Sec.~\ref{sec:conductance}, the accidental symmetry plays a crucial role in a flat band instability concerning a surface misorientation.

 Finally, the BdG Hamiltonian possesses spin-rotation symmetry, $[S_z, H(\bm{k})] = 0$ with $S_z= i s_z \tau_z$. As discussed below, the spin-rotation symmetry is essential for the stability of the surface flat bands in the even $\nu$ case.
\section{Topological numbers and surface states}
\label{sec:top}
\begin{figure*}[tbp]
\centering
 \includegraphics[width=18cm]{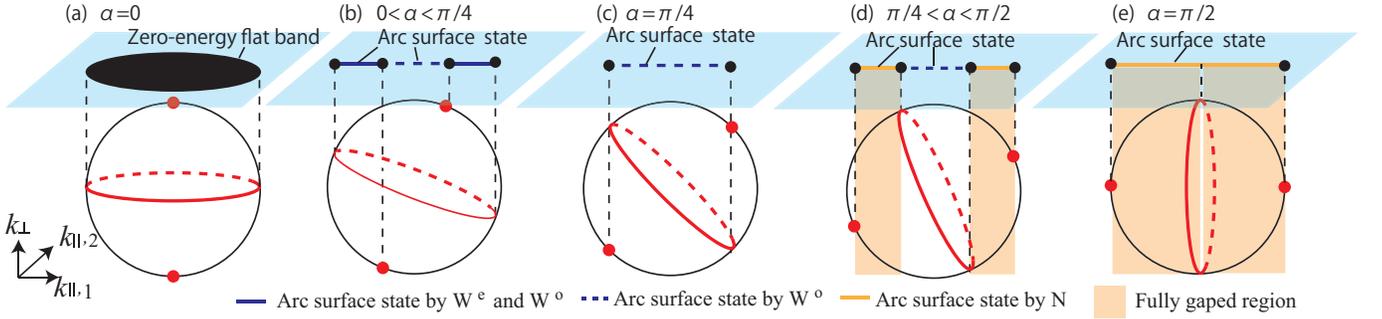}
  \caption{(color online) Relation between the misorientation angle $\alpha$ and the topological surface states in 3D chiral SCs with a spherical Fermi surface and $\nu \neq 0$. In the angle $0 < \alpha < \pi/4$, the zero-energy flat band suddenly disappears and instead an arc surface state appears according to the 1D winding number defined by Eq.~(\ref{eq:w-ky0}). When $\alpha$ is greater than $\pi/4$, fully gapped regions exist in the bulk BZ, which allows to exist a nontrivial TKNN number $N(k_{\parallel,1})$ defined by Eq.~(\ref{eq:N2-1}). The TKNN number ensures the existence of $2\nu$ arc surface states, where $2$ comes from the spin degeneracy. } \label{fig:chiral-angle}
 \end{figure*}
 Topological properties of nodal SCs in 3D are attributed to topological numbers of point and line nodes.  
 To characterize the topological numbers in 3D chiral SCs in which point and line nodes exist in the $k_z$ axis and on the $k_x k_y$ plane, we introduce 1D and 2D topological numbers associated with line and point nodes. These topological numbers are defined within lower dimensional subspaces in the BZ. Importantly, the 1D topological number originates from the interplay between topology and symmetry and is affected by the pseudo TRS and the spin-rotation symmetry, as we shall show in the following.  
 \subsection{One-dimensional topological number}
\label{subsec:1Dtop} 
To define the 1D topological number associated with a line node, we recall a winding number in TRI SCs (see Refs.~\onlinecite{Sato:2011,Schnyder:2011} for example). A key ingredient is a so-called chiral operator $\Gamma \equiv -i CT$ satisfying the anti-commutation relation $\{\Gamma , H(\bm{k})\} = 0$. Using the chiral operator and a BdG Hamiltonian, the 1D winding number is normally defined by
\begin{align}
 W (\bm{k}_{\parallel},\Gamma)\equiv \frac{-1}{4 \pi i}\int_{-\infty}^{\infty} d k_{\perp}\trace[\Gamma H^{-1}(\bm{k}) \partial_{k_{\perp}} H(\bm{k})], \label{eq:1dw}
\end{align}  
which takes an integer under the regularization~\cite{Sato:2011}. $\bm{k}_{\parallel}$ and $k_{\perp}$ are wave vectors parallel and perpendicular to a certain surface, respectively. Although 3D chiral SCs do not have TRS, they have the pseudo TRS (\ref{eq:pseudo-TRS}) instead. The combination of the pseudo TRS and PHS also satisfies the anticommutation relation with the BdG Hamiltonian, whereby leading to a chiral operator. Furthermore, the existence of inversion symmetry leads to the vanishing of the 1D winding number: $W(\bm{k}_{\parallel},\Gamma) =0 $ in the even $\nu$ case. (See Appendix~\ref{app:Blount} for more details). Hence, a line node in odd-parity SCs is generally unstable under TRS and inversion symmetry and a stable one requires an additional symmetry. In this paper, in order to acquire a stable line node, 
we use the spin-rotation symmetry provided in the previous section, which makes a line node stable~\cite{Kobayashi:2014}. In the following, we define a chiral operator and an associated 1D winding number for even and odd parity cases, respectively. 

In the even-parity case, a proper chiral operator is defined by $\Gamma_{\varphi_{\bm{k}}} \equiv U_{\varphi_{\bm{k}}}^{\dagger} \Gamma U_{\varphi_{\bm{k}}}$ and the corresponding 1D winding number is given by Eq.~(\ref{eq:1dw}) replacing $\Gamma$ with $\Gamma_{\varphi_{\bm{k}}}$:
\begin{align}
 W (\bm{k}_{\parallel} ,\Gamma_{\varphi_{\bm{k}}}) \equiv W^{\rm e} (\bm{k}_{\parallel}). \label{eq:w-even}
\end{align}
Since $\Gamma_{\varphi_{\bm{k}}}$ depends on $U_{\varphi_{\bm{k}}}$, the quantization of $W^{e}$ requires $ \partial_{k_{\perp}} \varphi_{\bm{k}} =0$, which is satisfied only for $k_{\perp} = k_z$ or $\varphi_{\bm{k}}=0$. For this reason, the topological surface state strongly depends on the relative angle between the line node and the surface. This is a key feature of 3D chiral SCs, resulting in instability of ZBCP in terms of the surface misorientation. 

On the other hand, using the local gauge transformation and the spin-rotation symmetry, the chiral operator in the odd-parity case is defined as $\Gamma_{\varphi_{\bm{k}}}^{\rm s} \equiv U_{\varphi_{\bm{k}}}^{\dagger} S_z CT U_{\varphi_{\bm{k}}}$. 
With the chiral operator $\Gamma_{\varphi_{\bm{k}}}^{\rm s}$, we define the 1D topological number as 
  \begin{align}
   W (\bm{k}_{\parallel} ,\Gamma_{\varphi_{\bm{k}}}^{\rm s}) \equiv W^{\rm o} (\bm{k}_{\parallel}). \label{eq:w-odd}
  \end{align}  
Here $\nu=0$ satisfies $\varphi_{\bm{k}}=0$. Thus, $ W^{\rm o} $ with $\nu=0$ is well-defined at any surface except for the surface being normal to the line node. On the other hand, $W^{\rm o}$ with $\nu \neq 0$ is quantized only for $k_{\perp} = k_z$ or $\varphi_{\bm{k}}=0$ due to the momentum dependence of the chiral operator. As a result, line-node-induced surface states will be sensitive not only to a surface misorientation, but also to the RSOI. 

 
Assuming that a pairing interaction is weak, $\Delta(\bm{k})$ is negligibly small far way from the Fermi surface. The weak pairing assumption allows us to simplify the calculation of the 1D winding numbers because the main contribution comes from momentum around the Fermi surface. Substituting Eqs. (\ref{eq:BdG}) and (\ref{eq:gapfn}) into Eqs. (\ref{eq:w-even}) and (\ref{eq:w-odd}), the 1D winding numbers are rewritten in a simple form~\cite{Sato:2011,Tanaka:2012,Schnyder:2012}. In $p_z$-wave SCs ($\nu=0$), it yields for any $k_{\perp}$
\begin{align}
 W^{\rm o} (\bm{k}_{\parallel}) =  \sum_{\mathcal{E} (\bm{k}) = 0} \sgn [\partial_{k_{\perp}} \mathcal{E} (\bm{k})] \cdot \sgn [k_z], \label{eq:w-pz}
\end{align}
where the summation is taken for $k_{\perp}$ satisfying $\mathcal{E} (\bm{k}) =0$. Whereas the 1D winding numbers with $\nu \neq 0$ are described as follows. When $k_{\perp} =k_z$,
\begin{align}
 W^{\rm e (o)} (k_x,k_y) =  \sum_{\mathcal{E} (\bm{k}) = 0} \sgn [\partial_{k_z} \mathcal{E} (\bm{k})] \cdot \sgn \left[k_z \right], \label{eq:w-kz}
\end{align}
and when $k_{\perp} \neq k_z$ and $k_y =0$,
\begin{align}
 W^{\rm e (o)} (k_{\parallel,1}) =  \sum_{\mathcal{E} (\bm{k}) = 0} \sgn [\partial_{k_{\perp}} \mathcal{E} (\bm{k})] \cdot \sgn [k_zk_x^{\nu}], \, \nu: {\rm odd \, (even)}. \label{eq:w-ky0}
\end{align}
Here $k_{\perp}$ and $k_{\parallel,1}$ lie in the $k_x k_z $ plane and are orthogonal to each other. To see the 1D winding numbers concretely, we shall consider a spherical Fermi surface $\mathcal{E} (\bm{k}) = \frac{\hbar^2}{2m} (\bm{k}^2-k_F^2)$, where $m$ is the mass of electron, and introduce the misorientation angle between the line node and the surface as $\alpha $ satisfying 
\begin{align}
 \left( \begin{array}{@{\,} c @{\,}} k_{\parallel,1} \\ k_{\perp}  \end{array} \right) = \begin{pmatrix} \cos \alpha & \sin \alpha \\ - \sin \alpha & \cos \alpha \end{pmatrix} \left( \begin{array}{@{\,} c @{\,}} k_{x} \\ k_{z}  \end{array} \right).
\end{align}
  In this case, Eq.~(\ref{eq:w-kz}), i.e., $\alpha =0$ leads to $2$ at the inside of the Fermi surface in the surface BZ. On the other hand, as we change the surface misorientation ($\alpha \neq 0$),  Eq.~(\ref{eq:w-ky0}) describes the surface state and is evaluated as follows. When $0< \alpha < \pi/4$, 
\begin{subequations}
\begin{align}
 W^{\rm e} (k_{\parallel,1})& = \begin{cases} -2 & -\cos \alpha < k_{\parallel,1} /k_F < - \sin \alpha, \\
                             2 & \sin \alpha < k_{\parallel,1} /k_F < \cos \alpha, \\
                          0 & \text{otherwise}, \end{cases} \\
 W^{\rm o} (k_{\parallel,1})& = \begin{cases} 2 &|k_{\parallel,1} /k_F| < \cos \alpha, \\
                          0 & \text{otherwise}, \end{cases}
 \end{align} \label{eq:w-ky0-1}
 \\
 \end{subequations}
 and when $\pi/4 < \alpha < \pi/2$, 
\begin{subequations}
\begin{align}
  W^{\rm e} (k_{\parallel,1})& = \begin{cases} -2 & -\sin \alpha < k_{\parallel,1} /k_F < - \cos \alpha, \\
                             2 & \cos \alpha < k_{\parallel,1} /k_F < \sin \alpha, \\
                          0 & \text{otherwise}, \end{cases} \\
 W^{\rm o} (k_{\parallel,1})& = \begin{cases} 2 & |k_{\parallel,1} /k_F| < \cos \alpha, \\
                          0 & \text{otherwise}, \end{cases}
 \end{align} \label{eq:w-ky0-2}
 \\
\end{subequations}
where $2$ is due to the spin degeneracy.
Equations (\ref{eq:w-ky0-1}) and (\ref{eq:w-ky0-2}) show a nontrivial winding number along the direction $k_y=0$ for a proper angle.
  \subsection{Two-dimensional topological number}
\label{subsec:2Dtop}    
  Due to the presence of point nodes ($\nu \neq 0$), there exists the Thouless-Kohmoto-Nightingale-den Nijs (TKNN) number~\cite{Thouless:1982} defined at a fixed $k_{\parallel,1}$, which also characterizes the topological structure:
  \begin{align}
   N(k_{\parallel,1}) = \frac{i}{2 \pi} \sum_{n \in occ} \int_{\rm BZ} d k_{\perp} d k_{\parallel,2} \epsilon^{ab} \partial_{k_a} \langle u_n (\bm{k}) | \partial_{k_b}| u_n (\bm{k}) \rangle, \label{eq:N2-1}
  \end{align}
 where $| u_n(\bm{k}) \rangle$ is an eigenstate of $H(\bm{k})$ and the summation is taken over all of occupied states. Here $k_{\parallel,2}$ is a direction orthogonal to $k_{\perp}$ and $k_{\parallel,1}$; $k_a$ and $k_b$ take $k_{\perp}$ and $k_{\parallel,2}$. Using eigenvectors of Eq.~(\ref{eq:BdG}), we obtain for both even and odd parity pair potentials
 \begin{align}
   N(k_{\parallel,1}) = \frac{2 \nu }{4 \pi} \int_{- \infty}^{\infty} \int_{- \infty}^{\infty}  d k_{\perp} d k_{\parallel,2}  \epsilon^{ab} \sin \theta_{\bm{k}} \partial_{k_a} \theta_{\bm{k}} \partial_{k_b} \varphi_{\bm{k}}, \label{eq:N2-2}
  \end{align} 
with $\theta_{\bm{k}} \equiv \tan^{-1}[\Delta_0 k_z (k_x^2+k_y^2)^{\frac{{\nu}}{2}}/ (\mathcal{E}(\bm{k})k_F^{\nu+1})]$. As a concrete model, let us consider a spherical Fermi surface given by $\epsilon (\bm{k}) = \frac{\hbar^2}{2m} \left[ \bm{k}^2-k_F^2 + \delta \left( k_{\perp}^2+k_{\parallel,2}^2 \right)^j \right]$ ($2j > \nu$), where the last term is added so as to satisfy the regularization at $k_{\perp}, k_{\parallel,2}  \to \infty$ and $\delta$ is an infinitesimal coefficient. When $k_{\parallel,1} = k_z$, i.e., $\alpha = \pi/2$, Eq.~(\ref{eq:N2-2}) is readily calculated as
\begin{align}
 N(k_z) = - \nu (1 + \sgn (k_F^2 -k_z^2)) \label{eq:N2-chiral}
\end{align} 
except for $k_z =0$ (see Fig.~\ref{fig:chiral-angle} (e)). When $k_{\parallel,1} \neq k_z$, $N(k_{\parallel,1})$ takes $- 2 \nu$ if $H(\bm{k})$ is fully gapped at a fixed $k_{\parallel,1}$. This condition is satisfied for $-\sin \alpha < k_{\parallel,1} /k_F < -\cos \alpha$ and $\cos \alpha < k_{\parallel,1} /k_F < \sin \alpha$ within $\pi/4 < \alpha < \pi/2$  (see Fig.~\ref{fig:chiral-angle} (d)).
   
 \subsection{Topological surface states}
\label{subsec:top-surface}
 The 1D and 2D topological numbers ensure the existence of zero-energy surface states via the bulk-boundary correspondence. We shall consider the semi-infinite superconductor on $x_{\perp} > 0$, where $x_{\perp}$ is the conjugate coordinate of $k_{\perp}$ and the surface is at $x_{\perp} =0$. The BdG equation is given by
 \begin{align}
  H(x_{\perp},\bm{k}_{\parallel}) \Psi (x_{\perp},\bm{k}_{\parallel}) =  E(\bm{k}_{\parallel}) \Psi (x_{\perp},\bm{k}_{\parallel}), \label{eq:BdGeq}
 \end{align}
 with the boundary condition $\Psi (0,\bm{k}_{\parallel}) =0$. The zero-energy surface state satisfies $E(\bm{k}_{\parallel})=0$. We first discuss a zero-energy flat band associated with a line node. If a chiral symmetry $\Gamma$ exists in $H(\bm{x}_{\perp},\bm{k}_{\parallel})$, it demands that the number of positive energy states is equal to that of negative energy states for $E(\bm{k}_{\parallel}) \neq 0$ due to the anti-commutation relation between $\Gamma$ and $H$. Thus, as we define the number of zero-energy states with $\Gamma = +(-)$ as $N^{+}_0 $ ($N^-_0$), stable zero energy states exist only when $N^+_0 \neq N^-_0$. It is for the reason that chiral symmetry preserving perturbations shift the surface states from zero energy within a pair of the zero-energy states with opposite chirality.  The number of zero-energy states indeed connects with the winding number (\ref{eq:1dw}):~\cite{Sato:2011}
 \begin{align}
  W(\bm{k}_{\parallel},\Gamma) = N^-_0 - N^+_0. \label{eq:index}
 \end{align}
 Since the pseudo TRS plays the same role as TRS at the specific momentum such as $k_{\perp} = k_z$ and $k_y=0$, we can apply Eq.~(\ref{eq:index}) to Eqs.~(\ref{eq:w-even}) and (\ref{eq:w-odd}) there. In the case of the spherical Fermi surface, a zero energy flat band will appear at the inside of the Fermi surface in the surface BZ when the surface is parallel to the line node ($\alpha =0$) (see Fig.~\ref{fig:chiral-angle} (a)). In contrast to the zero-energy flat band, as we shift the misorientation angle from $0$, the many zero-energy states vanish and instead arc surface states emerge in the line $k_y=0$ as remnants of the zero-energy flat band according to Eqs. (\ref{eq:w-ky0-1}) and (\ref{eq:w-ky0-2}). (See Fig.~\ref{fig:chiral-angle} (b)-(d)). The vanishing of the zero-energy flat band contrasts sharply with that in TRI SCs and is one of the main results in this paper.
 
 Similarly to a line node, a pair of point nodes induces an arc surface state terminating the projected point nodes on the surface BZ (see Fig.~\ref{fig:chiral-angle} (e)). Let us consider the semi-infinite superconductor ($x>0$) with the spherical Fermi surface. From Eq.~(\ref{eq:N2-chiral}), the TKNN number takes $-2 \nu$ within $|k_z| <k_F$. Hence, $2 \nu$ arc surface states with the chiral linear dispersion appears in the projected surface BZ. The same arc surface state has been studied in chiral SCs without the line node, such as superfluid helium A phase~~\cite{Volovik:2003,Buchholtz:1981,Hara:1986} and Sr$_2$RuO$_4$~\cite{Yamashiro:1997,Yamashiro:1998,Honerkamp:1998,Kashiwaya:2011,Yada:2014}.
  It would be noted that 1D and 2D topological number induced surface states coexist in the angle $\pi/4 < \alpha<\pi/2$, together with the emergence of fully gapped regions in the bulk BZ. (See Fig.~\ref{fig:chiral-angle} (d)).

\section{Tunneling conductance}
\label{sec:conductance}
\begin{figure}[tbp]
\centering
 \includegraphics[width=8cm]{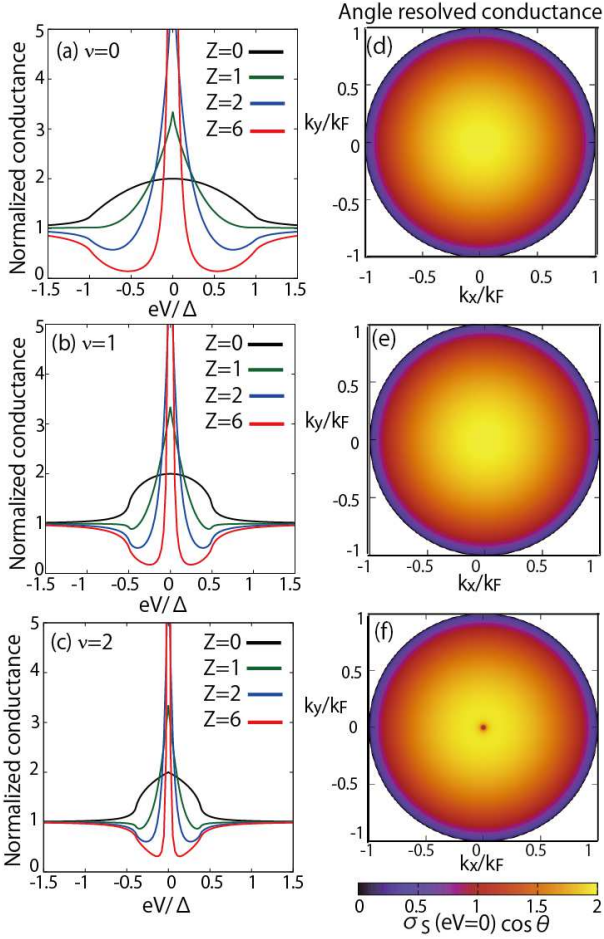}
  \caption{(color online) Normalized conductances for $\nu=0$ (a), $\nu=1$ (b), and $\nu=2$ (c) with various barrier potentials $Z=0, 1,2,$ and $6$. Angle resolved conductances $\sigma_S(eV=0,\theta,\phi) \cos \theta$ with $Z=6$ for $n=0,1,$ and 2 are shown in the pictures (d), (e), and (f), respectively, which indicate the formation of the zero-energy flat band at (001) plane as the function of $(k_x,k_y)$.} \label{fig:conductance}
 \end{figure}
To test these expectations in the topological argument, we discuss in the following tunneling conductance in a 3D normal metal/insulator/chiral superconductor (N/I/S) junction. It is well-known that a surface zero-energy flat band induces a sharp ZBCP for line nodal SCs such as high-$T_c$ cuprates and noncentrosymmeric SCs. Normally, the ZBCP plays an important role to distinguish a nodal SC from a fully gapped one and is robust against TRS preserving perturbations, e.g., a nonmagnetic impurity. 

 In contrast, as discussed in the previous section, the ZBCP in 3D chiral SCs is expected to be more fragile than that in TRI SCs. This is because the chiral operator depends on the pseudo TRS and/or the spin-rotation symmetry. Hence, the breaking of lattice symmetry or spin-rotation symmetry will shift the flat band from zero energy. The purpose of this section is to demonstrate these fragilities of the ZBCP under the effect of the surface misorientation and the surface RSOI. These perturbations preserve TRS, but breaks each of the accidental symmetries. To see these phenomena, we shall consider an N/I/S junction with a flat interface perpendicular to the z-axis and the spherical Fermi surface, i.e., $\mathcal{E} (\bm{k}) = \frac{\hbar^2 \bm{k}^2}{2m} - \mu$, where $\mu$ is the chemical potential. In this setup, we calculate tunneling conductance using the Blounder-Thinkham-Klapwijk (BTK) formula~\cite{Blonder:1982}, with taking into account the effect of the surface misorientation and the surface RSOI. By solving the BdG equation (\ref{eq:BdGeq}) for $E(\bm{k}_{\parallel})=eV$ and using the quasi-classical approximation, i.e., $\mu \gg |E(\bm{k}_{\parallel})|, |\Delta (\bm{k})|$, the wave function ansatz for the normal state ($z <0$) and the superconducting state ($z >0$) is given by
\begin{subequations}
\begin{align}
\Psi^{\rm N}_{\sigma}(z,\bm{k_{\parallel}})  &= \psi^{\rm N}_{e,\sigma}e^{i \bm{k} \cdot \bm{r}}+ \sum_{\sigma'=\pm} \Big( a_{\sigma,\sigma'} \psi^{\rm N}_{h, \sigma' }e^{i \bm{k} \cdot \bm{r}} \notag \\
&+ b_{\sigma, \sigma'} \psi^{\rm N}_{e, \sigma'}e^{i \tilde{\bm{k}} \cdot \bm{r}} \Big) \ \ z <0, \\
\Psi^{\rm S}(z,\bm{k_{\parallel}}) & =  \sum_{\sigma'=\pm} \left( c_{\sigma'} \psi^{\rm S}_{e, \sigma' }e^{i \bm{k} \cdot \bm{r}} + d_{\sigma'} \psi^{\rm S}_{h, \sigma'}e^{i \tilde{\bm{k}} \cdot \bm{r}} \right) \ \ z >0,
\end{align}
\end{subequations}
with 
\begin{subequations}
\begin{align}
&\psi^{\rm N}_{e,\sigma} = 1/2 [1+\sigma,1-\sigma,0,0]^T, \\
&\psi^{\rm N}_{h,\sigma} = 1/2 [0,0,1+\sigma,1-\sigma]^T, \\
&\psi^{\rm S}_{e,\sigma} = 1/2 [1+\sigma,1-\sigma,(1-\sigma) \eta \Gamma_+,(1+\sigma) \Gamma_+]^T, \\
&\psi^{\rm S}_{h,\sigma} = 1/2 [(1+\sigma)\Gamma_-,(1-\sigma) \eta \Gamma_-,1-\sigma, 1+\sigma]^T,
\end{align}
\end{subequations}
where $\Gamma_+ (\theta ,\phi)= \Delta^{\ast} (\theta ,\phi)/[E + \sqrt{E^2 - |\Delta (\theta ,\phi)|^2}$, $\Gamma_- (\theta ,\phi)= \Delta (\pi - \theta ,\phi)/[E + \sqrt{E^2 - |\Delta (\pi - \theta ,\phi)|^2}]$ and $\eta=-(+)$ for the even (odd) parity pairing. The superscripts $N$ and $S$ indicate the normal and superconducting states and the subscripts $e$ and $h$ describe electron and hole states, respectively. $\tilde{\bm{k}} = (k_x,k_y-k_z)$ and the all trajectories of electron and hole have the same incident angle $(k_x,k_y,k_z)=k_F (\sin \theta \cos \phi, \sin \theta \sin \phi, \cos \theta)$ since we assume that the chemical potential for the normal state is the same as that for the superconducting state. 
The coefficients $a_{\sigma,\sigma'}$, $b_{\sigma,\sigma'}$, $c_{\sigma}$, and $d_{\sigma}$ are determined by the boundary conditions:
\begin{subequations}
\begin{align}
&\Psi^{\rm N}_{\sigma} (0_-,\bm{k_{\parallel}}) = \Psi^{\rm S} (0_+,\bm{k_{\parallel}}),\\ 
&\frac{d \Psi^{\rm S}}{dz}\Big|_{z=0_+}  - \frac{d \Psi^{\rm N}_{\sigma}}{dz}\Big|_{z=0_-} = \frac{2 m U_0}{ \hbar^2} \Psi^{\rm S} (0_+,\bm{k_{\parallel}}) , 
\end{align} \label{eq:boudary-cond}
\\
\end{subequations}
where the insulating barrier at $z=0$ is modeled as the $\delta$-function $V(z) = U_0 \delta(z)$. Solving Eq. (\ref{eq:boudary-cond}), we obtain the normal reflection and Andreev reflection coefficients $b_{\sigma,\sigma'}$ and $a_{\sigma,\sigma'}$, which give the transmissivity in the N/I/S junction as~\cite{Kashiwaya:2000}
\begin{align}
 \sigma_{\rm S} (eV, \theta, \phi)&= 1+ \frac{1}{2} \sum_{\sigma,\sigma'=\pm} [ |a_{\sigma,\sigma'}|^2 - |b_{\sigma,\sigma'}|^2 ] \notag \\
 & = \sigma_{\rm N} \frac{1+ \sigma_{\rm N} |\Gamma_+|^2 + (\sigma_{\rm N}-1) |\Gamma_+|^2 |\Gamma_-|^2}{|1+(\sigma_{\rm N} -1) \Gamma_+ \Gamma_- |^2}, \label{eq:transmissivity}
\end{align}
where $\sigma_{\rm N} = \frac{4}{4+Z_0^2}$ and $Z_0= \frac{2mU_0}{\hbar^2 k_F \cos \theta} \equiv \frac{Z}{\cos \theta}$. Integrating $\sigma_{\rm S}(eV, \theta,\phi)$ with respect to the all incident angles of injected electrons, the normalized tunneling conductance is given by 
\begin{align}
 \sigma(eV) = \frac{\int_0^{2\pi} d \phi \int_0^{\pi/2} d \theta \, \sigma_S (eV, \theta, \phi) \sin \theta \cos \theta }{\int_0^{2\pi} d \phi \int_0^{\pi/2} d \theta \, \sigma_N \sin \theta \cos \theta}. \label{eq:conductance}
\end{align} 
We numerically calculate the normalized conductance (\ref{eq:conductance}) and the transmissivity (\ref{eq:transmissivity}) for $\Delta (\theta, \phi)= \Delta_0 \cos \theta$ ($\nu=0$), $\Delta_0 \cos \theta \sin \theta e^{i \phi}$ ($\nu=1$), and $\Delta_0 \cos \theta \sin \theta e^{i 2 \phi}$ ($\nu=2$) with various barrier potentials, which are shown in Fig.~\ref{fig:conductance}. Due to the line node at $k_z =0$, the zero-energy flat band appears at the (001) plane in all cases with $Z=6$ (see Fig.~\ref{fig:conductance} (d), (e), and (f)) and we obtain a sharp ZBCP under the high barrier potential (see Fig.~\ref{fig:conductance} (a), (b), and (c)). 

\subsection{Effect of surface misorientation}
\begin{figure*}[tbp]
\centering
 \includegraphics[width=16cm]{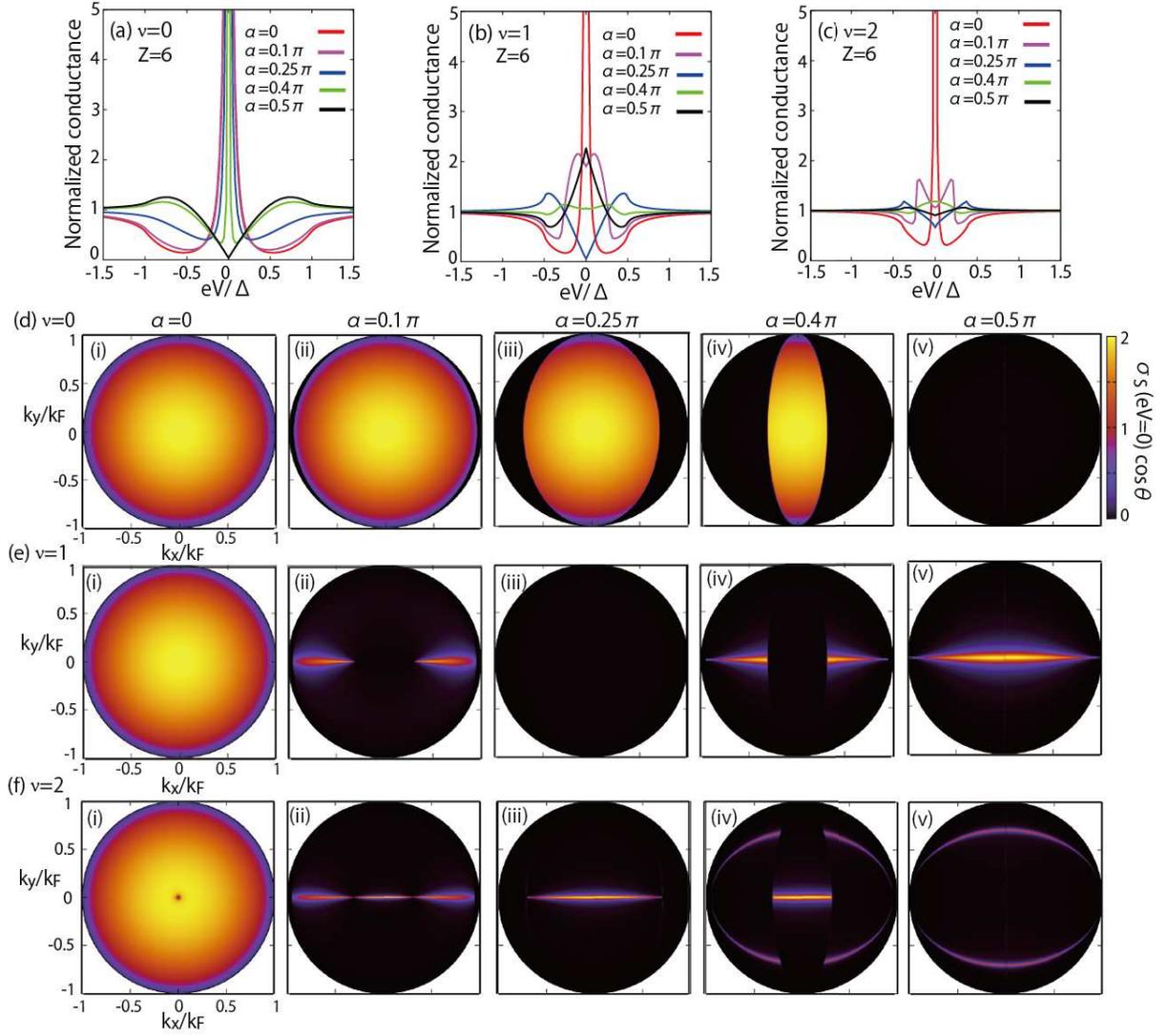}
  \caption{(color online) Normalized conductances with the barrier potential $Z=6$ and various misorientation angles $\alpha = 0, 0.1\pi , 0.25 \pi , 0.4 \pi$ and $0.5 \pi$ for $n=0$ (a), $n=1$ (b), and $n=2$ (c). Figures (d), (e), and (f) show each angle resolved conductance $\sigma_S (eV=0,\theta,\phi) \cos \theta$ as varying the misorientation angle $\alpha$.} \label{fig:misorientation}
 \end{figure*}
\label{subsec:misorientation}
To show the zero-energy flat band instability in terms of the misorientation angle, we rotate the coordinate of the pair potential in the $k_z k_x$ plane instead of rotating the surface: $(k_x ,k_y ,k_z) \to (k_x \cos \alpha - k_z \sin \alpha, k_y , k_x \sin \alpha + k_z \cos \alpha ) $, where $\alpha$ is the misorientation angle defined in  Sec.~\ref{subsec:1Dtop}. The pair potential for $\nu =0,1$ and $2$ become
\begin{align}
 \Delta (\theta, \phi , \alpha) &= \Delta_0 ( \sin \theta \cos \phi \sin \alpha +\cos \theta \cos \alpha), \label{eq:delta0-alpha} \\
\Delta (\theta, \phi , \alpha) &= \Delta_0 ( \sin \theta \cos \phi \sin \alpha +\cos \theta \cos \alpha) \notag \\
 \times & ( \sin \theta \cos \phi \cos \alpha - \cos \theta \sin \alpha + i \sin \theta \sin \phi),  \label{eq:delta1-alpha} \\
 \Delta (\theta, \phi , \alpha) &= \Delta_0 ( \sin \theta \cos \phi \sin \alpha +\cos \theta \cos \alpha) \notag \\
 \times & ( \sin \theta \cos \phi \cos \alpha - \cos \theta \sin \alpha + i \sin \theta \sin \phi)^2.  \label{eq:delta2-alpha}
 \end{align}
Substituting Eqs.~(\ref{eq:delta0-alpha}), (\ref{eq:delta1-alpha}), and (\ref{eq:delta2-alpha}) into Eqs.~(\ref{eq:transmissivity}) and (\ref{eq:conductance}), we numerically calculate the normalized conductance and the angle resolved conductance for each case, where $\alpha$ takes $0, 0.1 \pi, 0.25 \pi, 0.4\pi$, and $0.5 \pi$. The resultant conductance is shown in Fig.~\ref{fig:misorientation}. When $\alpha =0$, the sharp ZBCP and the surface zero-energy flat band appear in all cases due to the presence of the line node. As slightly changing $\alpha$, in the $p$-wave SC ($\nu=0$), the ZBCP remains and the zero-energy flat band still forms at the interior of the projected line node as shown in Fig.~\ref{fig:misorientation} (d-ii). In contrast, the ZBCP and the surface zero-enrgy flat band in the chiral SCs ($\nu=1$ and $2$) vanish abruptly (see Figs.~\ref{fig:misorientation} (e-ii) and (f-ii)). The reason for the disappearance of the flat band is understood from the definition of the 1D wingding number (\ref{eq:w-kz}) and (\ref{eq:w-ky0}). Hence, the 1D winding number for $\nu=1$ and $2$ is quantized only when $\partial_{k_{\perp}} \varphi_{\bm{k}}=0$, while that for $\nu=0$ is well-defined unless $k_{\perp}= k_{x}$. Therefore, the ZBCP in 3D chiral SCs ($\nu \neq 0$) is fragile against the surface misorientation and clearly different from the behavior of the ZBCP in the $p$-wave SC. Furthermore, in $\nu=1$ and $\nu=2$, an arc surface state emerges for the misorientation angle $0 < \alpha \le \pi/2$ due to the presence of the point nodes. When $0<\alpha < \pi/4$, the arc surface state originates from the 1D winding number~(\ref{eq:w-ky0}). According to~Eq.~(\ref{eq:w-ky0-1}), the arc surface state in Fig.~\ref{fig:misorientation} (e-ii) is induced by $W^{\rm e} $, while that in Fig.~\ref{fig:misorientation} (f-ii) comes from $W^{\rm o}$. 
As $\alpha$ reaches $\pi/4$, the projected positions of the line and point nodes overlap, so that the zero-energy states vanish for $\nu=1$ and that remains for $\nu=2$ as shown in Fig~\ref{fig:misorientation} (e-iii) and (f-iii). When $ \pi/4 < \alpha < \pi/2$, the TKNN number $N$ becomes nontrivial because the pair potential is fully gaped in $k_y k_z$ plane within $-\sin \alpha < k_x /k_F < -\cos \alpha$ and $\cos \alpha < k_x /k_F < \sin \alpha$. The TKNN number ensures the existence of an arc surface state for $\nu=1$ (see Fig.~\ref{fig:misorientation} (e-iv)) and double one for $\nu=2$ (see Fig.~\ref{fig:misorientation} (f-iv)). Interestingly, a mixed surface state emerges in Fig.~\ref{fig:misorientation} (f-iv) with $\nu=2$, in which the 1D winding number $W^{\rm o}$ induced surface state exists at $|k_x /k_F| < \cos \alpha$ and the TKNN number $N$ induced one at $- \cos \alpha<k_x/k_F < - \sin \alpha$ and $\sin \alpha < k_x /k_F < \cos \alpha$. Finally, achieving $\alpha = \pi/2$, the TKNN number induced arc surface state only exists.   

\subsection{Effect of Rashba spin-orbit interaction}
\label{subsec:Rashba}
\begin{figure}[tbp]
\centering
 \includegraphics[width=8cm]{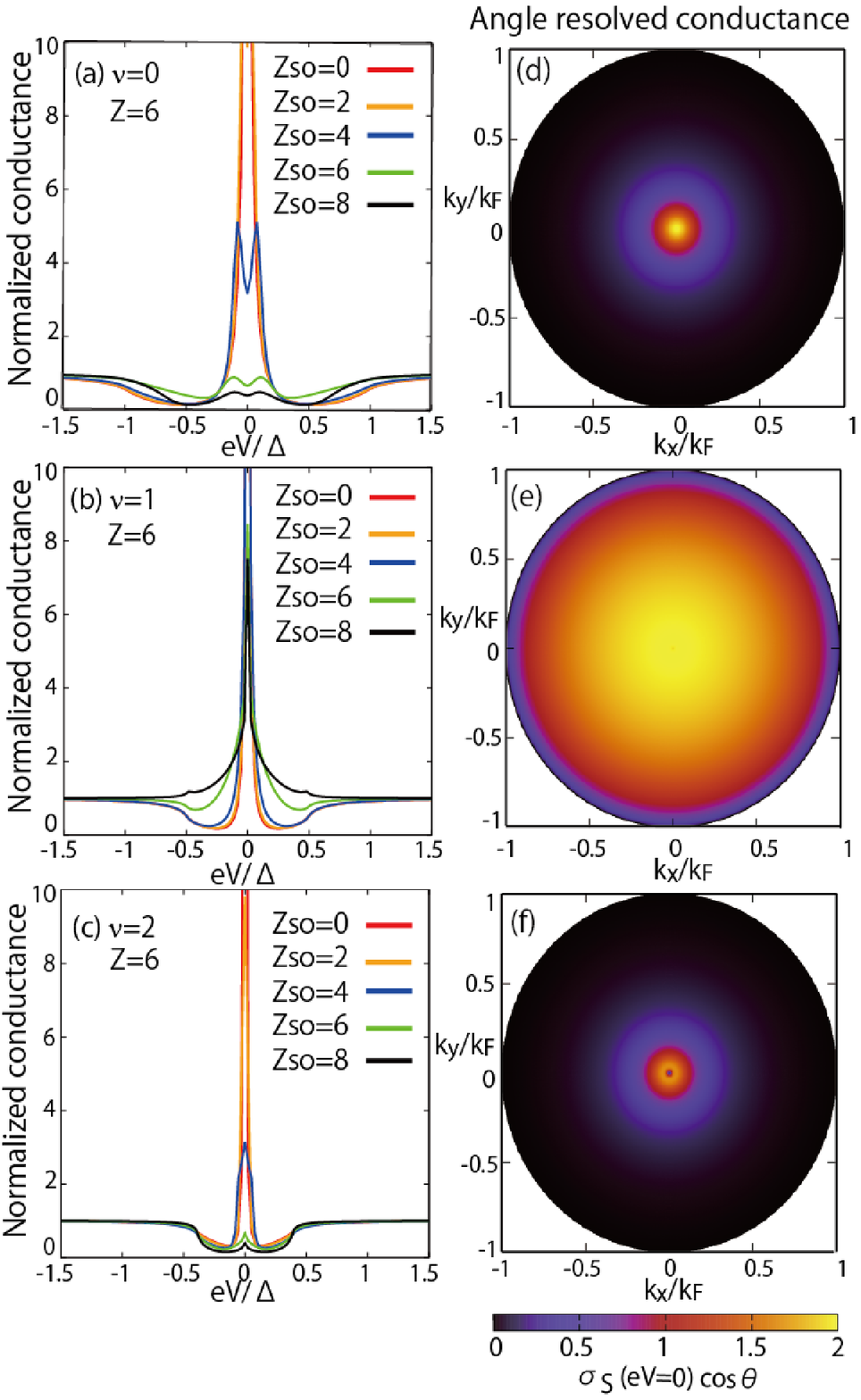}
  \caption{(color online) Normalized conductances for $n=0$ (a), $n=1$ (b), and $n=2$ (c) with the barrier potential $Z=6$, taking into account the RSOI $Z_{\rm SO} = 0,2,4,6$ and $8$. Angle resolved conductances $\sigma_S(eV=0,\theta,\phi) \cos \theta$ with $Z=6$ and $Z_{\rm SO}=8$ for $n=0,1,$ and 2 are shown in the pictures (d), (e), and (f), respectively. A large region of the zero-energy surface states disappears due to the RSOI when $\nu$ is even. } \label{fig:conductance3}
 \end{figure}

Next, we introduce the surface RSOI in order to investigate the flat band instability caused by the breaking of the spin-rotation symmetry. The interface at $z=0$ breaks inversion symmetry, which thus gives rise to the RSOI. We take into account this effect in the BTK formula based on Refs~\onlinecite{Wu:2010a,Wu:2010b}. Assuming that the RSOI is localized at $z=0$, we modify the insulating barrier potential as
\begin{align}
 V(z) = (U_0+ U_1 (\bm{s} \times \hat{\bm{k}}) \cdot \hat{\bm{z}}) \delta (z),
\end{align} 
where $\hat{\bm{k}}$ and $\hat{\bm{z}}$ mean a unit vector of $\bm{k}$ and $\bm{z}$, respectively. It follows the modified boundary condition:
\begin{subequations}
\begin{align}
&\Psi^{\rm N}_{\sigma} (0_-,\bm{k_{\parallel}}) = \Psi^{\rm S} (0_+,\bm{k_{\parallel}}),\\ 
&\frac{d \Psi^{\rm S}}{dz}\Big|_{z=0_+}  - \frac{d \Psi^{\rm N}_{\sigma}}{dz}\Big|_{z=0_-} \notag 
\\ &= \frac{2 m}{ \hbar^2} \begin{pmatrix} U_0 & U_{\rm SO} & 0 & 0 \\ U_{\rm SO}^{\ast} & U_0 & 0 & 0 \\ 0&0& U_0 & -U_{\rm SO}^{\ast}  \\ 0&0&  -U_{\rm SO} & U_0 \end{pmatrix} \Psi^{\rm S}|_{z=0_+} , 
\end{align} \label{eq:boudary-cond-SOC}
\\
\end{subequations}
where $U_{\rm SO} = i U_1 \sin \theta e^{-i \phi}$. By solving the BdG equation under the boundary condition (\ref{eq:boudary-cond-SOC}), $a_{\sigma,\sigma'}$ and $b_{\sigma,\sigma'}$ obey the simultaneous linear equations: 
\begin{align}
 &\begin{pmatrix} i Z_{1} & 2 -i Z_0 & i \Gamma_+ & i \Gamma_+ Z_{1} \\ 
                    2-i Z_0  & iZ_{1}^{\ast} & i \eta \Gamma_+ Z_{1}^{\ast} & i \eta \Gamma_+ Z_0 \\
                    i \Gamma_- Z_{1} & - i \Gamma_- Z_0 & 2+iZ_0 & iZ_{1} \\
                    i \eta \Gamma_- Z_0 & -i \eta \Gamma_- Z_{1}^{\ast} & -i Z_{1}^{\ast} & - (2+iZ_0)  \end{pmatrix}
  \left( \begin{array}{@{\,}c @{\,}} a_{\sigma, +} \\ a_{\sigma,-} \\ b_{\sigma,+} \\ b_{\sigma,-} \end{array} \right) \notag \\
  &= \begin{cases} \left( \begin{array}{@{\,}c @{\,}} \Gamma_+ (2-i Z_0) \\ -i \eta \Gamma_+ Z_{1}^{\ast} \\ -iZ_0 \\ i Z_{1}^{\ast} \end{array} \right)  &\text{if } \sigma=+, \\
  \left(\begin{array}{@{\,}c @{\,}} -i \Gamma_+ Z_{1} \\ \eta \Gamma_+ (2-iZ_0) \\ - iZ_{1} \\ iZ_0 \end{array}\right) &\text{if } \sigma=- ,\end{cases} \label{eq:coeff-Rashba}
\end{align}
where $Z_{1} = \frac{i 2 mU_{1} \sin \theta e^{-i \phi}}{\hbar^2 k_F \cos \theta} \equiv Z_{\rm SO} i e^{-i\phi} \frac{\sin \theta}{\cos \theta}$. Also, the normal conductance is given by 
\begin{align}
 \sigma_N = \frac{4(4 + |Z_{1}|^2 + Z_0^2)}{(4+|Z_{1}|^2-Z_0^2)^2  +16 Z_0^2}.
\end{align}
Numerically solving Eq.~(\ref{eq:coeff-Rashba}), we determine the transmissivity and the normalized conductance. The calculated normalized conductance and angle resolved conductance are shown in Fig.~\ref{fig:conductance3}. As expected, Figs.~\ref{fig:conductance3} (a) and (c) show the suppression of the ZBCP in the normalized conductance with increasing the magnitude of the RSOI, while Fig.~\ref{fig:conductance3} (b) maintains the ZBCP under the RSOI. Along with the suppression of the ZBCP, the disappearance of the zero-energy flat band is verified in Fig.~\ref{fig:conductance3} (d) and (f) via the angle resolved conductance in $eV=0$. Thus, the obtained results agree well with the topological arguments. Hence, in even $\nu$, the RSOI breaks the spin-rotation symmetry and leads to the flat band instability, while the suppression of the ZBCP does not occur in odd $\nu$ because the pseudo TRS only stabilize the flat band and is still preserved under the RSOI. 
 
\section{Summary}
\label{sec:sum}
 \begin{table*}[tbp]
\centering
\caption{Stability of surface flat bands and types of the chiral operators for several superconducting phases with a line node. The second column shows the definition of the chiral operators for each phase, where $\mathcal{S}$ ($\mathcal{S}^2=-1$) is an additional symmetry such as spin-rotation symmetry and mirror-reflection symmetry. The third and fourth columns describe the influence of the surface misorientation and the surface RSOI on the surface zero-energy flat bands, respectively. Here $\checkmark$ ($\times$) indicates (un)stable flat bands.} \label{tab:sum}
\begin{tabular}{cccc}
\hline \hline
Systems & Chiral operator &  Misorientation & RSOI \\ \hline
TRI noncentrosymmetric SCs~\cite{Yada:2011,Brydon:2011,Schnyder:2012} & $-iCT$ & $\checkmark$ & $\checkmark$ \\ 
TRI even-parity SCs~\cite{Kashiwaya:2000,Lofwander:2001,Wu:2010b} & $-iCT$ & $\checkmark$ &  $\checkmark$ \\
TRI odd-parity SCs & $\mathcal{S}CT$ & $\checkmark$ & $\times$ \\
Chiral even-parity SCs & $-iU_{\varphi_{\bm{k}}}^{\dagger}CTU_{\varphi_{\bm{k}}}$ & $\times$ &  $\checkmark$ \\
Chiral odd-parity SCs & $U_{\varphi_{\bm{k}}}^{\dagger}\mathcal{S}CTU_{\varphi_{\bm{k}}}$ & $\times$ &  $\times$ \\ 
\hline \hline
\end{tabular}

 \end{table*}
In summary, we have discussed surface states in 3D chiral SCs with $p_z (p_x +ip_y)^{\nu}$-wave pairing symmetry and found fragility of surface zero-energy flat bands against the surface misorientation and the surface RSOI. These instabilities are due to the breaking of the protecting symmetries: the pseudo TRS and the spin-rotation symmetry. Using the 1D winding number, we have shown that zero-energy flat bands in the 3D chiral SCs are protected by the accidental chiral symmetry consisting of the pseudo TRS and/or the spin-rotation symmetry. Thus, the resulting flat bands are more fragile than is normally understood. We have demonstrated the suppression of the ZBCP in the N/I/S junction by numerically calculating tunnel conductance, with taking into account the effect of the surface misorientation and the surface RSOI. As a result, we found the suppression of the ZBCP in terms of the surface misorientation in all of 3D chiral SCs ($\nu \neq 0$) because the pseudo TRS is sensitive to the breaking of lattice symmetry. Also, including the RSOI breaks the spin-rotation symmetry, resulting in the decrease of the ZBCP with increase in the magnitude of the RSOI in odd-parity SCs. We summarized the chiral operators discussed in this paper and the flat band instabilities in Table~\ref{tab:sum}, in which TRI noncentrosymmetric and TRI even-parity SCs are included. It is noteworthy that similar flat band instabilities associated with the breaking of chiral symmetry have been also discussed in 2D systems in the disordered limit~\cite{Atkinson:2000,Wimmer:2010,Queiroz:2014}. Proximity effect of 
the flat bands into diffusive normal metal is also an interesting problem~\cite{proximity1,proximity2,proximity3,proximity4,proximity5}. 


Finally, we mention the implication of our results for heavy fermion compounds UPt$_3$~\cite{Sauls:1994,Joynt:2002,Schemm:2014}, URu$_2$Si$_2$~\cite{Kasahara:2007,Shibauchi:2014,Schemm:2015}, and SrPtAs~\cite{Nishikubo:2011,Biswas:2013,Fischer:2014}. In this paper, we have mainly focused on TRS breaking SCs with $p_z (p_x +ip_y)^{\nu}$-wave pairing symmetry, assuming that the normal Hamiltonian possesses TRS. In the heavy fermion systems, however, the gap functions are representation of a point group and they can be more complicated. Nevertheless, as long as they take the form $p_z (p_x +ip_y)^{\nu}$ near the Fermi surface, our result is applicable even for heavy fermion SCs. 
In our calculation, we do not determine the spatial dependence 
of the pair potential~\cite{Hara:1986,Matsumoto:1995}. As far as we are considering zero energy states, the obtained results will not be changed even if the spatial depletion of the pair potential near the surface is taken into account~\cite{Kashiwaya:2000,Tanuma:2007,Mizushima:2014}. 

\section{Acknowledgments}
We thank K. Yada and A. Yamakage for valuable discussions. This work was supported in part by the “Topological Quantum Phenomena” Grant-in Aid for Scientific Research on Innovative Areas from the MEXT of Japan (No. 22103005), the “Topological Materials Science” Grant-in Aid for Scientific Research on Innovative Areas from the MEXT of Japan (No. 15H05853, 15H05855), a Grant-in-Aid for Scientific Research B (Grant No. 15H03686) (YT), a Grant-in-Aid for Challenging 
Exploratory Research (Grant No. 15K13498) (YT), a Grant-in-aid for JSPS Fellows (No. 256466) (SK) and a 
Grant-in-Aid for Scientific Research B (No. 25287085) (MS).

\appendix
\section{Vanishing of 1D winding number in TRI odd-parity superconductors}
\label{app:Blount}
In the presence of TRS and PHS, we always have the chiral operator $\Gamma = -i CT$, and then it is possible to define the 1D winding number $W(\bm{k}_{\parallel},\Gamma)$ by Eq.~(\ref{eq:1dw}). In what follows, we show that the 1D winding number $W$ vanishes in the case of an odd-parity pair potential owing to inversion symmetry.  We start with the general BdG Hamiltonian described by 
\begin{align}
 H = \frac{1}{2} \sum_{\bm{k},\alpha, \alpha'} \left( c_{\bm{k} \alpha}^{\dagger}, c_{-\bm{k} \alpha} \right) H(\bm{k}) \left( \begin{array}{@{\,} c @{\,}} c_{\bm{k} \alpha'} \\ c_{-\bm{k} \alpha'}^{\dagger} \end{array}\right) , \label{eq:BdG1}
\end{align}
where $H(\bm{k})$ is given by
\begin{align}
 H(\bm{k}) = \begin{pmatrix} \mathcal{E} (\bm{k})_{\alpha \alpha'} & \Delta (\bm{k})_{\alpha \alpha'} \\ \Delta (\bm{k})_{\alpha \alpha'}^{\dagger} & -\mathcal{E} (-\bm{k})_{\alpha \alpha'}^{T} \end{pmatrix}. \label{eq:BdG2}
\end{align}
$c_{\bm{k}\alpha}^{\dagger}$ ($c_{\bm{k}\alpha'}$) represents the
creation (annihilation) operator of an electron with momentum $\bm{k}$. The
suffix $\alpha$ represents other degrees of freedom such as spin, orbital, and
sublattice indices. $\epsilon (\bm{k})_{\alpha \alpha'}$ and
$\Delta (\bm{k})_{\alpha \alpha'}$ are the Hamiltonian in the normal state
and gap function, respectively. The TRI BdG Hamiltonian possesses TRS:
\begin{align}
 \Theta H(\bm{k}) \Theta^{-1} = H(-\bm{k})^{\ast}, \, \Theta = \begin{pmatrix} U_{\alpha \alpha} & 0 \\ 0 & U_{\alpha \alpha'}^{\ast} \end{pmatrix}, \label{eq:TRS}
\end{align}
where $U$ satisfies $U \mathcal{E}(\bm{k}) U^{\ast} = \mathcal{E} (-\bm{k})^{\ast}$ and $U \Delta (\bm{k}) U^T = \Delta (-\bm{k})^{\ast}$, and PHS:
\begin{align}
C H(\bm{k}) C^{-1} = - H(-\bm{k})^{\ast}, \, C = \begin{pmatrix} 0 & \delta_{\alpha \alpha'} \\ \delta_{\alpha \alpha'} & 0 \end{pmatrix}. \label{eq:PHS}
\end{align}
 Thus, the combination of Eqs.~(\ref{eq:PHS}) and (\ref{eq:TRS}) satisfies $\{CT,H(\bm{k})\}=0$, which gives the chiral operator
\begin{align}
\Gamma = -i CT = \begin{pmatrix} 0 & -i U^{\ast}_{\alpha \alpha'} \\ -i U_{\alpha \alpha'} & 0\end{pmatrix}.
\end{align}
For convenience, we choose the basis with the chiral operator being diagonal such that
\begin{align}
U_{\Gamma}^{\dagger} \Gamma U_{\Gamma} = \begin{pmatrix} \delta_{\alpha \alpha'} & 0 \\ 0 & -\delta_{\alpha \alpha'} \end{pmatrix}, \ \ U_{\Gamma} = \begin{pmatrix} \delta_{\alpha \alpha'} & -i \delta_{\alpha \alpha'} \\ -i U_{\alpha \alpha'} &U_{\alpha \alpha'} \end{pmatrix}. 
\end{align}
In this basis, $H (\bm{k})$ becomes off-diagonal form
\begin{align}
 U^{\dagger}_{\Gamma} H (\bm{k}) U_{\Gamma} = \begin{pmatrix} 0 & q (\bm{k}) \\  q(\bm{k})^{\dagger} & 0 \end{pmatrix}, \label{eq:H-Gamma}
\end{align}
with 
\begin{align}
 q(\bm{k}) \equiv  i \mathcal{E} (\bm{k}) + \Delta (\bm{k}) U.
\end{align}
With this basis, the 1D winding number is rewritten as~\cite{Sato:2011}
\begin{align}
 W(\bm{k}_{\parallel},\Gamma) &= \frac{i}{4 \pi} \int_{-\infty}^{\infty} d k_{\perp} \trace [ \Gamma H^{-1} (\bm{k}) \partial_{k_\perp} H (\bm{k})] \notag \\
  &= \frac{1}{ 2\pi} {\rm Im} \left[ \int_{-\infty}^{\infty} d k_{\perp} \partial_{k_{\perp}} \ln \det q(\bm{k}) \right],
\end{align}
where $k_{\perp}$ is momenta perpendicular to the surface. In the TRI odd-parity SC, the BdG Hamiltonian hosts inversion symmetry in addition to the chiral symmetry:
\begin{align}
 \tilde{P} H(\bm{k}) \tilde{P}^{-1} = H(-\bm{k}), \, \tilde{P} = \begin{pmatrix} P_{\alpha \alpha'}& 0 \\ 0 & - P_{\alpha \alpha'} \end{pmatrix}, \label{eq:IS}
 \end{align}
 where $P$ is a real matrix and satisfies $P^2=1$. Under the unitary transformation $U_{\Gamma}$, the inversion operator also transforms into
\begin{align}
 U_{\Gamma}^{\dagger} \tilde{P} U_{\Gamma} = \begin{pmatrix} 0 & -i P_{\alpha \alpha'} \\ i P_{\alpha \alpha'} &0\end{pmatrix}, \label{eq:IS-Gamma}
\end{align}
where $[U,P]=0$ is assumed because $P$ always acts trivially on the spin space. From Eqs.~(\ref{eq:H-Gamma}) and (\ref{eq:IS-Gamma}), the addition constraint is added on $q(\bm{k})$:
\begin{align}
 P q^{\dagger}(\bm{k}) P = - q (-\bm{k}). \label{eq:q-IS}
\end{align}
Furthermore, $U$ and $q(\bm{k})$ satisfies the relation:
\begin{align}
 U q (\bm{k}) U^{\dagger} = q(-\bm{k})^T. \label{eq:q-TRS}
\end{align}
Combining Eqs. (\ref{eq:q-IS}) and (\ref{eq:q-TRS}), we obtain
\begin{align}
 U P q(\bm{k})^{\dagger}P U^{\dagger} = -q (\bm{k})^T. \label{eq:q-TP}
\end{align}
Finally, we take the determinant for the both side of Eq.(\ref{eq:q-TP}),
\begin{align}
 \det q(\bm{k})^{\dagger} &= \det (-1) \det q(\bm{k})^T \notag \\
                               &= \det q(\bm{k})^T,
\end{align}
where $\det (-1)=1$ due to the spin degrees of freedom. Therefore, $\det q(\bm{k})$ is a real function of $\bm{k}$, which immediately proves $W(\bm{k}_{\parallel},\Gamma)=0$ for any $\bm{k}_{\parallel}$.

\end{document}